\documentclass{sig-alternate}

\usepackage{amsmath}
\usepackage{multirow}
\usepackage{algorithmic}
\usepackage{url}
\usepackage{color}

\makeatletter
\newif\if@restonecol
\makeatother

\newcommand{\ignore}[1]{}

\begin{document}

\title{Dynamic Memory Allocation Policies for\\Postings in Real-Time Twitter Search}

\numberofauthors{3}
\author{
Nima Asadi$^{1}$, Jimmy Lin$^{1}$, and Michael Busch$^{2}$\\[1ex]
\affaddr{$^{1}$ University of Maryland, College Park \qquad $^{2}$ Twitter}\\
\email{nima@cs.umd.edu, jimmylin@umd.edu, @michibusch}
}

\maketitle
\begin{abstract}
We explore a real-time Twitter search application
where tweets are arriving at a rate of several
thousands per second. Real-time search demands that
they be indexed and searchable immediately, which leads to a number
of implementation challenges. In this paper, we focus on one aspect:\
dynamic postings allocation policies for index structures
that are completely held in main memory. The core issue can be
characterized as a ``Goldilocks Problem''. Because memory remains
today a scare resource, an allocation policy that is too aggressive
leads to inefficient utilization, while a policy that is too
conservative is slow and leads to fragmented
postings lists. We present a dynamic postings allocation policy that
allocates memory in increasingly-larger ``slices'' from a small number of large,
fixed pools of memory. Through analytical models and experiments, we
explore different settings that balance time (query
evaluation speed) and space (memory utilization).
\end{abstract}

\section{Introduction}

The rise of social media and other forms of user-generated content
challenges the traditional notion of search as operating on either
static documents collections or document collections that evolve
slowly enough where periodically running a batch indexer (e.g.,
every hour) suffices. We focus on real-time search in the context of
Twitter:\ users demand to know
what's happening {\it right now}, especially in response to breaking
news stories and other shared events such as hurricanes in the northeastern
United States, the death of prominent figures, or televised political debates.
For this, they often
turn to real-time search.

The context of this study is Twitter's Earlybird retrieval
engine~\cite{Busch_etal_ICDE2012}, which serves over two billion
queries a day with an average query latency of 50 ms. Usually, tweets
are searchable within 10 seconds after creation (most of the latency
is from the processing pipeline---tweet indexing itself takes less
than a millisecond). The service as a whole is of course a complex,
distributed system with many components. In this paper, we focus on
one aspect---dynamic memory allocation policies for postings
allocation.

A key feature of Earlybird is that it incrementally indexes tweets as
they are posted and makes them immediately searchable. The
indexing process naturally requires allocating space for postings in a
dynamic manner---we adopt a zero-copy approach that yields
non-contiguous postings lists.  The fundamental challenge boils down
to a ``Goldilocks Problem'', since memory today remains a scarce
resource. A policy that is too aggressive in allocating memory for
postings leads to inefficient utilization, because much of the
allocated space will be empty. On the other hand,
a policy that is too conservative slows the system, since memory
allocation is a relatively costly operation and postings lists will
become fragmented.  Ideally, we'd like to strike a balance between the
two extreme and find a ``sweet spot'' that balances speed with utilization.

We present a dynamic postings allocation policy that allocates
increasingly-larger ``slices'' from a small number of memory pools.
The production system, which we previously described in Busch
et al.~\cite{Busch_etal_ICDE2012}, deploys a particular instantiation
of a general framework, which we articulate for the first time
here. Until now, we have not thoroughly explored alternative parameter
settings in a rigorous and controlled manner. Thus, the contribution
of this paper is a detailed study of the design space for dynamic
postings allocation in the context of our basic
framework:\ we present both an analytical model for estimating time
and space costs, which is subsequently validated by experiments on
real data.

\section{Operational Requirements}
\label{section:bg}

To set the stage, we begin by discussing differences and similarities between
real-time search and  ``traditional''
(e.g., web) search. First, two similarities:

\begin{list}{\labelitemi}{\leftmargin=1em}
\vspace{-0.2cm}
\setlength{\itemsep}{-2pt}

\item {\it Low-latency, high-throughput query evaluation.} Users are
  impatient and demand results quickly.

\item {\it In-memory indexes.} The only practical way to achieve
  necessary performance requirements is to maintain all index
  structures in memory.

\vspace{-0.2cm}
\end{list}

\noindent There are important
differences as well:

\begin{list}{\labelitemi}{\leftmargin=1em}
\vspace{-0.2cm}
\setlength{\itemsep}{-2pt}

\item {\it Immediate data availability.} In real-time search, documents
  arrive rapidly, and users expect content to be searchable within seconds.
  This means that the indexer must achieve both low
  latency and high throughput. This requirement departs from common
  assumptions that indexing can be considered a batch
  operation. Although web crawlers achieve high throughput, it
  is generally not expected that crawled content be indexed
  immediately---an indexing delay of
  minutes to hours may be acceptable. This allows efficient indexing with
  batch processing frameworks such as
  MapReduce~\cite{Dean_Ghemawat_OSDI2004}. In contrast, real-time search demands
  that documents be searchable in {\it seconds}.

\item {\it Shared mutable state.} A real-time search engine must
  handle shared mutable state in a multi-threaded execution
  environment with concurrent indexing and retrieval operations. In
  contrast, concurrency-related challenges are simpler to handle in
  web search:\ for example, it is possible to atomically ``swap out''
  an old index with an updated new index without service disruption. Such
  a design would be impractical in real-time search.

\item {\it Dominance of the temporal signal.} The nature of real-time
  search means that temporal signals are important for relevance ranking.
  This contrasts with web search, where document
  timestamps have a relatively minor role in determining
  relevance (news search being the obvious exception).
  This holds implications for how postings should be organized in index
  structures.

\end{list}

\section{Baseline Architecture}

Twitter's production real-time search service is a complex
distributed system spanning many machines, the details of which are
beyond the scope of this paper. In this study, we specifically focus
on Earlybird, which is the core retrieval engine.
For the purposes of this paper, Earlybird receives boolean queries
and returns tweets that satisfy the query, sorted in reverse
chronological order. No relevance scoring is performed, which is,
functionally speaking, handled by another component. Incoming tweets
are hash partitioned across a number of replicated Earlybird instances, so
that each individual instance serves a fraction of all tweets.

To understand our contributions, it is necessary to first provide some technical
background. Here, we summarize material presented in a previous
paper~\cite{Busch_etal_ICDE2012}, but refer the reader to the original
source for details.

\subsection{Earlybird Overview}
\label{section:baseline:overview}

Earlybird is built on top of the open-source Lucene search
engine\footnote{http://lucene.apache.org/} and adapted to meet the
demands of real-time search discussed in
Section~\ref{section:bg}. The system is written completely
in Java, primarily for three reasons:\ to take advantage of the
existing Lucene Java codebase, to fit into Twitter's
JVM-centric development environment, and to take advantage of the
easy-to-understand memory model for concurrency offered by Java and
the JVM. Although this decision poses inherent challenges in terms of
performance, with careful engineering and memory management we believe
it is possible to build systems that are comparable in performance to
those written in C/C++.

As with nearly all modern retrieval engines, Earlybird maintains an inverted
index:\ postings are maintained in forward chronological order (most
recent last) but are traversed {\it backwards} (most recent first);
this is accomplished by maintaining a pointer to the current end of
each postings list (more details in the next section).

Earlybird supports a full boolean query language consisting of conjunctions
(ANDs), disjunctions (ORs), negations (NOTs), and phrase
queries. Results are returned in reverse chronological order, i.e.,
most recent first. Boolean query evaluation is relatively
straightforward, and in fact we use Lucene query operators ``out of
the box'', e.g., conjunctive queries correspond to postings
intersections, disjunctive queries correspond to unions, and phrase
queries correspond to intersections with positional
constraints. Lucene provides an abstraction for postings lists and
traversing postings---we provide an implementation for our custom
indexes, and are able to reuse existing Lucene query evaluation code.

A particularly noteworthy aspect of Earlybird is the manner in which it
handles shared mutable state (concurrent index reads and writes) using
lightweight memory barriers. As this is not germane to the subject of
this paper, we refer the reader elsewhere~\cite{Busch_etal_ICDE2012} for details.
However, it is worth mentioning that the general strategy for
handling concurrency is to limit the scope of data structures that hold
shared mutable state. This is accomplished as follows:\ each instance
of Earlybird manages multiple index segments (currently 12), and each segment
holds a relatively small number of tweets (currently, $2^{23} \sim
8.4$ million tweets). Ingested tweets first fill up a segment before
proceeding to the next one. Therefore, at any given time, there is at
most one index segment actively being modified, whereas the remaining
segments are read-only. Once an index segment ceases to accept new
tweets, we can convert it from a write-friendly structure into an
optimized and compressed read-only structure.

Due to this design, our paper is only concerned with the active index
segment within an Earlybird instance:\ only for that index do we need to
allocate memory for postings dynamically. This is described in more
detail next.

\subsection{Active Index Segment}
\label{section:baseline:active}

As we argued in Section~\ref{section:bg}, the dominance of the temporal signal
is a major distinguishing characteristic of real-time search, compared
to traditional (web) search. The implication of this is that it would
be desirable to traverse postings in reverse temporal order for query
evaluation. Although this is not an absolute requirement, such a
traversal order is the most convenient.

Following this reasoning further, it appears that existing approaches
to index structure organization are not appropriate. The information retrieval literature
discusses two types of indexes:\ document sorted and frequency/impact
sorted. The latter seems unsuited for real-time search. What about
document-sorted indexes? If we assign document ids to new tweets in
ascending order, there are two obvious possibilities when indexing
new documents:

First, we could append new postings to the ends of postings
lists. However, this would require us to read postings {\it backwards}
to achieve a reverse chronological traversal order. Unfortunately,
this is not directly compatible with modern index compression
techniques. Typically, document ids are converted into document gaps,
or differences between consecutive document ids. These gaps are then
compressed with integer coding techniques such as $\gamma$ codes, Rice
codes, or PForDelta~\cite{Yan_etal_WWW2009,ZhangJiangong_etal_WWW2008}. It would
be tricky for gap-based compression to support backwards
traversal. Prefix-free codes ($\gamma$ and Rice codes) are meant to be
decoded only in the forward direction. More recent techniques such as
PForDelta are block-based, in that they code relatively large blocks
of integers (e.g., 128 document ids) at a time. Reconciling this with
the desire to have low-latency indexing would require additional
complexity, although none of these issues are technically
insurmountable.

Alternatively, we could prepend new postings to the beginnings of
postings lists. This would allow us to read postings in the forward
direction and preserve a reverse chronological traversal
order. However, this presents memory management challenges, i.e., how
would space for new postings be allocated? We are unaware of any work
that has explored this strategy. Note that
the na\"ive implementation using linked lists would be hopelessly
inefficient:\ linked list traversal is slow
due to the lack of reference locality and
predictable memory access patterns. Furthermore, linked lists have rather
large memory footprints due to object overhead and the need to store
``next'' pointers.

Based on the above analysis, it does not appear that
real-time search capabilities can be efficiently realized with obvious
extensions or adaptations of existing techniques.

Earlybird implements the following solution:\ each posting is simply
a 32-bit integer---24 bits are devoted to storing the document id and
8 bits for the term position. Since tweets are limited to 140
characters, 8 bits are sufficient to hold term positions.\footnote{\small If
  a term appears in the tweet multiple times, it will be represented
  with multiple postings.}  Therefore, a list of postings is simply an
integer array, and indexing new documents involves inserting elements
into a pre-allocated array. Postings traversal in reverse
chronological order corresponds to iterating through the array
backwards. This organization also allows every array position to be a
possible entry point for postings traversal to evaluate queries. In
addition, it allows for binary search (to find a particular document
id), and doesn't require any additional skip-pointers~\cite{Moffat_Zobel_1996} to
enable faster traversal through the postings lists. Finally, this
organization is cache friendly, since array traversal involves linear
memory scans and this predictable access pattern provides
prefetch cues to the hardware.

In essence, the design punts on the problem of postings
compression---but we feel that this is a reasonable design choice
given its simplicity and the above advantages. Furthermore, since the
active index segment holds relatively few tweets, a particular segment
doesn't spend much time in the uncompressed state. Once an index
segment stops accepting new tweets, it is converted into an optimized
read-only structure:\ we apply a variant of PForDelta after reversing the order of
the postings.

Having provided adequate background, we finally arrive at the heart of
this paper:\ the allocation of space for postings lists. Obviously,
this process needs to be dynamic, since postings list growth is only
bounded by the size of the collection itself. There are a few
challenges to overcome:\ postings lists vary significantly in size,
since term and document frequencies are Zipfian (roughly). As a result, it is
tricky to choose the correct amount of memory to allocate for each
term's postings (i.e., size of the integer array). Selecting a value
that is too large leads to inefficient memory utilization, because
most of the allocated space for storing postings will be empty. On the
other hand, selecting a value that is too small leads to waste:\ time,
obviously, for memory allocation (which is a relatively costly operation), but also space because
non-contiguous postings require pointers to chain together (in the
limit, allocating one posting at a time is akin to a linked
list). Furthermore, during postings traversal, blocks that are too small
result in suboptimal memory access patterns (e.g., due to cache
misses, lack of memory prefetching, etc.). This is exactly the
``Goldilocks Problem'' we described in the introduction.

\begin{figure}[t]
\hspace{-0.1cm}\centering\includegraphics[width=1.0\linewidth]{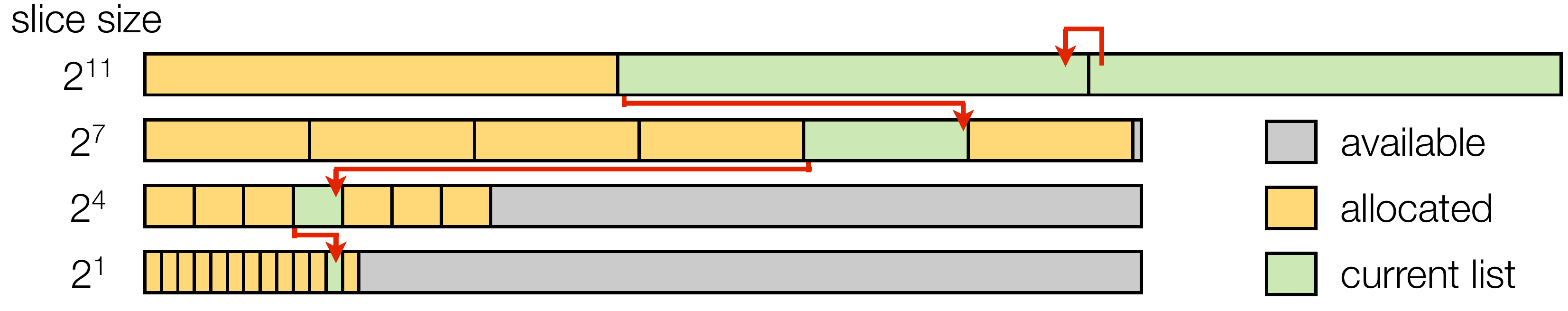}
\caption{Organization of the active index segment where
  tweets are ingested. Increasingly larger slices are
  allocated in the pools to hold postings. Except for slices in pool 1
  (the bottom pool), the first 32 bits are used for storing the
  pointer that links the slices together. Pool 4 (the top pool) can
  hold multiple slices for a term. The green rectangles illustrate the
  the ``current'' postings list that is being written into.}
\vspace{-0.25cm}
\label{figure:postings-slices}
\end{figure}

Our approach to address these issues is to create four separate
``pools'' for holding postings. Conceptually, each pool can be treated
as an unbounded integer array. In practice, pools are large integer
arrays allocated in $2^{15}$ element blocks; that is, if a pool fills
up, another block is allocated, growing the pool. In each pool, we
allocate ``slices'', which hold individual postings belonging to a
term. In each pool, the slice sizes are fixed:\ they are $2^1$, $2^4$,
$2^7$, and $2^{11}$, respectively (see
Figure~\ref{figure:postings-slices}). For convenience, we will refer
to these as pools 1 through 4, respectively. When a term is first
encountered, a $2^1$ integer slice is allocated in the first pool,
which is sufficient to hold postings for the first two term
occurrences. When the first slice runs out of space, another slice of $2^4$
integers is allocated in pool 2 to hold the next $2^4-1$ term
occurrences (32 bits are used to serve as the ``previous'' pointer,
discussed below).  After running out of space, slices are allocated in
pool 3 to store the next $2^7-1$ term occurrences and finally
$2^{11}-1$ term occurrences in pool 4. Additional space is allocated
in pool 4 in $2^{11}$ integer blocks as needed.

One advantage of this strategy is that no array copies are required as
postings lists grow in length---which means that there is no garbage to
collect. However, the tradeoff is that postings are non-contiguous and
we need a mechanism to link the
slices together.
Addressing slice positions is accomplished using 32-bit pointers:\ 2
bits are used to address the pool, 19--29 bits are used to address the
slice index, and 1--11 bits are used to address the offset within the
slice. This creates a symmetry in that postings and addressing
pointers both fit in a standard 32-bit integer. The dictionary
maintains pointers to the current ``tail'' of the postings list using
this addressing scheme (thereby marking where the next posting should
be inserted and where query evaluation should begin). Pointers in the same format are used to
``link'' the slices in different pools together and, possibly,
multiple slices in pool 4. In all but the first pool, the first 32
bits of each slice are used to store this ``previous'' pointer.

To conclude this section, we provide some performance figures, summarized from~\cite{Busch_etal_ICDE2012}.  The
basic configuration of an Earlybird server is a commodity machine with two
quad-core processors and 72 GB memory. A fully-loaded active index
segment with 16 million documents occupies about 6.7 GB memory. On
such a segment, we achieve 17000 queries per second with a 95th percentile latency of
$<$100 ms and 99th percentile latency of $<$200 ms using 8 searcher
threads. In a stress test, we evaluated Earlybird indexing performance under
near 100\% CPU utilization. We achieve 7000 tweets per second (TPS) indexing rate at 95th
percentile latency of 150 ms and 99th percentile latency of 180 ms.
Indexing latency is relatively insensitive to tweet arrival
rate; at 1000 TPS we observe roughly the same latencies as at 7000
TPS.

\subsection{Generalizing the Solution}
\label{section:baseline:solution}

It is evident that Earlybird
represents a specific instantiation of a general solution to the
problem of dynamically allocating postings for real-time search:\ from
a small number of large memory pools, we allocate increasingly larger
slices for postings as more term occurrences are encountered. Within
this general framework, a particular instantiation can be described by
$Z = \langle z_0, z_1, ..., z_{P-1}\rangle$, the slice size settings (as powers of two),
where $P$ is the number of pools. For example, in the production
deployment, $Z = \langle 1,4,7,11 \rangle$. For best utilization of
bits in addressing pointers, it is helpful to restrict $|P|$ to a power
of two also.

Note that this framework provides a general solution to real-time
indexing (not only tweets):\ we simply assume that slices hold spaces for postings and
pointers to previous slices. In the case of tweets, both postings
and pointers are 32-bit integers, but nothing in our model precludes
other encodings. Thus, for the remainder of this paper, we measure
postings in terms of ``memory slots''. For simplicity, we
assume that pointers also fit in a memory slot, but if this isn't the
case, a small constant factor adjustment will suffice.

How ``optimal'' is the current production deployment, compared to
alternative configurations? Prior to this study, we have not attempted
to answer this question in a rigorous, controlled
fashion. In this paper, we tackle this question as follows:\ First, we
define a cost model in terms of speed and memory usage, the two
characteristics we seek to balance. Second, we develop an analytical
model that allows us to assess the time and space costs of a
particular configuration. Finally, for promising configurations
identified by our analytical model, we follow up with
experiments.

\section{Data}

Since our analytical model makes use of real data to estimate
parameters, we begin by describing our datasets.
For tweets, we used the Tweets2011 corpus created for the TREC 2011
microblog track.\footnote{\small
  \url{http://trec.nist.gov/data/tweets/}} The corpus is comprised of
approximately 16 million tweets over a period of two weeks (24th
January 2011 until 8th February, inclusive) which covers both the time
period of the Egyptian revolution and the US Superbowl. Different
types of tweets are present, including replies and retweets. The
corpus represents a sample of the entire tweet stream, but since
tweets are hash partitioned across multiple Earlybird instances in
production, experiments on these tweets is a reasonably accurate
facsimile of studying an individual Earlybird instance. Even though we have
access to all tweets, we purposely conducted experiments on this
publicly available collection so that others will be able to replicate our results.

Three different sets of queries were used in our evaluation.  First,
we took the TREC 2005 terabyte track ``efficiency''
queries\footnote{\small
  \url{http://www-nlpir.nist.gov/projects/terabyte/}} (50,000 queries
total).  Second, we sampled 100,000 queries randomly from the AOL
query log~\cite{Pass_InfoScale_2006}, which contains around 10 million
queries in total. Our sample preserves the original query length
distribution.  Finally, we used queries from the TREC 2011 microblog
track. However, since there were only 50 queries (which is
insufficient for efficiency experiments), we augmented the queries by
first generating the power set of all query terms and then used the
``related queries'' API of a commercial search engine to harvest query
variants. In this way, we were able to construct a set of
approximately 3100 queries.

Our choice of these three datasets represented an attempt
to balance several factors. Although we have access to actual Twitter
query logs, experiments on them would have several drawbacks:\ First,
due to their proprietary nature, our results would not be
replicable. Second, the majority
of Twitter queries are trending hashtags (or queries containing
trending hashtags), which are not particularly interesting from
an efficiency point of view (similar to
head navigational queries in web search). Furthermore, we'd like
to study the types of information needs that real-time search {\it
  could} solve, not exactly what the service is doing right
now. Thus, triangulating based on three query sets paints a more
complete picture:\ the AOL queries represent general web queries;
the TREC efficiency queries are representative of
{\it ad hoc} queries, closer to the ``torso'' of the query
distribution (mostly informational, as opposed to navigational);
finally, the TREC microblog queries represent a forward-looking
conception of what real-time search might evolve into
(at least according to retired intelligence analysts at NIST).
Finally, all three of our datasets are available to
researchers (we intend to release our expanded microblog queries).

\section{Analytical Model}

Given a collection of documents $\mathcal{C}$ and a set of queries $\mathcal{Q}$, we
define a cost function for memory usage. The total memory ``wasted''
is equal to the memory allocated for postings minus the size of the
postings list (i.e., number of postings), summed across all terms $t$ in the collection:
\begin{displaymath}
\sum_{t \in \mathcal{C}}{\left[\textrm{Memory}(t; Z) - \textrm{Size}(\textrm{Postings}(t))\right]}
\end{displaymath}
Since the size of postings is constant for a given collection,
we can simply define the
memory cost as follows (which we'd like to minimize):
\begin{equation}
\mathcal{C}_M = \sum_{t \in \mathcal{C}}{\textrm{Memory}(t; Z)}
\label{equation:memorycost}
\end{equation}

Similarly, let us define the time cost as the time it would take to
read all postings (end to end) for all query terms in each query of
$\mathcal{Q}$.
\begin{equation}
\mathcal{C}_T = \sum_{Q \in \mathcal{Q}}\sum_{q \in Q}{\textrm{TimeToRead}(\textrm{Postings}(q))}
\label{equation:timecost}
\end{equation}
Note that this cost function does not actually take into account time
spent in query evaluation (e.g., intersection of postings lists for
conjunctive query processing). We decided to factor out those costs for two 
reasons:\ First, to
support a simpler model (since a large number of postings traversal
techniques are available, each with different optimizations and
tradeoffs). Second, even if we wished to, it is unclear how we could analytically
model postings intersection time, which is a function of term occurrences in
real-world data.

The advantage of our model is that instantiating it with parameters is fairly easy.
If we assume that term
frequencies in a collection follow a Zipfian distribution (a standard assumption 
in information retrieval), we can analytically estimate the
memory cost for various $Z$ configurations.
Similarly, if the postings length distribution of query terms is known, we can
analytically model the time cost as well. With models of the two costs,
we can find configurations that strike a desired memory/speed balance.
The remainder of this section explains how we accomplish this.

\subsection{Memory Cost Estimation}
\label{section:memorycost}

Given that the frequency of a term $t$ in a collection is $f_r$, 
and the pool settings is $Z = \langle z_0, z_1, ..., z_{P-1}\rangle$,
we can calculate the exact number of memory slots
required to hold the postings list of term $t$.
Let us define a step function $\mathcal{M}$ that maps a frequency to the number of memory slots
required by configuration $Z$. First, we recursively define a set of
thresholds $\theta_i$'s on the frequencies as follows:\footnote{Note that the maximum frequency for a term is bounded and therefore the set
of $\theta_i$'s is a finite set.}
\begin{displaymath}
\theta_i = \begin{cases}
2^{z_0}, & i = 0\\
\theta_{i - 1} + (2^{z_i} - 1), & 0 < i \leq P\\
\theta_P + (i - P)\times (2^{z_{P - 1}} - 1), & i > P\\
\end{cases}
\end{displaymath}
For each term frequency interval $\{f_r \in \mathbb{N}\, |\, \theta_{i - 1} < f_r \leq \theta_i\}$
the value of the step function $\mathcal{M}$ can be computed as follows:
\begin{displaymath}
\mathcal{M}(f_r) =
\begin{cases}
\theta_0, & f_r \leq \theta_0\\
\theta_i + i, & \theta_{i - 1} < f_r \leq \theta_i \; (i > 0)\\
\end{cases}
\end{displaymath}
This function computes the amount of memory (i.e., number of slots)
that needs to be allocated to store pointers along with the actual
postings. Given function $\mathcal{M}$, we can rewrite Equation~(\ref{equation:memorycost}) as:
\begin{equation}
\mathcal{C}_M = \sum\limits_{1 \leq t \leq |V|}{\mathcal{M}(f_r(t))}
\label{equation:memorycost:direct}
\end{equation}
\noindent where $f_r(t)$ is the frequency of term $t$, and $|V|$ is the size of the vocabulary.
Making a standard simplifying assumption, if we rank the terms in the collection with respect to their frequencies, the resulting
pairs of $\langle r, \bar{f}_r \rangle$ (where $\bar{f}_r$ is normalized) form a Zipfian distribution, with
the following probability mass function (PMF):
\begin{equation}
p(x) = {x^{-\alpha} \over H_{|V|, \alpha} }
\label{equation:zipfian:pmf}
\end{equation}
\noindent where $H_{\rho, \alpha}$ is the $\rho_{th}$ generalized harmonic number, and $\alpha$
is a parameter. From Equation~(\ref{equation:zipfian:pmf}), one can estimate a term
frequency given the rank of term $r(t)$ and the total number of terms in the collection $N$ as:
\begin{displaymath}
f_r(t) = N \times p(r(t))
\end{displaymath}
Thus, we can rewrite Equation~(\ref{equation:memorycost:direct}) as follows:
\begin{equation}
\mathcal{C}_M = \sum\limits_{1 \leq r \leq |V|}{\mathcal{M}(N \times p(r))}
\label{equation:memorycost:zipfian}
\end{equation}
\noindent where $r$ is the rank (with respect to frequency) of a term in the collection.
Equation~(\ref{equation:memorycost:zipfian}) gives an analytical model for estimating
the memory cost of indexing a particular collection, given $N$ (total number
of terms) and the characteristic Zipfian parameter $\alpha$.

Furthermore, we can speed up the computation of
Equation~(\ref{equation:memorycost:zipfian}) by exploiting the
fact that the PMF of a Zipfian distribution is a one-to-one function. In this way, based
on the definition of the step function $\mathcal{M}$, we have:
\begin{align*}
& \theta_{i - 1} < N \times p(r) \leq \theta_i \Rightarrow\\
& \theta_{i - 1} < {{N \times r^{-\alpha}}\over{H_{|V|, \alpha}}} \leq \theta_i \Rightarrow\\
& \left(\theta_{i-1} \times {H_{|V|, \alpha} \over N}\right) < r^{-\alpha} \leq \left(\theta_i \times {H_{|V|, \alpha} \over N}\right) \Rightarrow\\
& {\theta_{i - 1}^{-1\over\alpha} \times \underbrace{\left({H_{|V|, \alpha} \over N}\right)^{-1\over\alpha}}_{\text{$\beta$}}} >
r \ge {\theta_i^{-1\over\alpha} \left({H_{|V|, \alpha} \over N}\right)^{-1\over\alpha}} \Rightarrow\\
& \beta \theta_{i - 1}^{-1\over\alpha} > r \geq \beta \theta_{i}^{-1\over\alpha}
\end{align*}
Therefore, we can rewrite Equation~(\ref{equation:memorycost:zipfian}) as follows by substituting the
above in the definition of $\mathcal{M}$:
\begin{align*}
\mathcal{C}_{M} &= \sum\limits_{1 \leq r \leq |V|}{\mathcal{M}(N \times p(r))}\\
&= \sum\limits_{|V| \ge r \ge {\beta \times \theta_0^{-1\over\alpha}}} {\theta_0}
+ \sum\limits_{\theta_k \neq \theta_0}{ \sum\limits_{\beta \theta_{k - 1}^{-1\over\alpha} < r \leq {\beta \theta_k^{-1\over\alpha}}}
{(\theta_k + k)}}\\
&= \left( |V| - \beta \theta_0^{-1\over\alpha} + 1\right) \theta_0
+ \sum\limits_{\theta_k \neq \theta_0}{\beta \left(\theta_{k - 1}^{-1\over\alpha} - \theta_k^{-1\over\alpha} \right) (\theta_k + k)}
\end{align*}

To summarize, given a characteristic Zipfian parameter $\alpha$, the total number of terms $N$,
and a configuration $Z$, we can compute the memory cost of indexing a particular collection in closed form.

\subsection{Time Cost Estimation}
\label{section:timecost}

We now turn to our analytical model of time cost, that of the sum of reading postings
lists corresponding to all query terms.
Let us assume that the cost of reading postings for a configuration $Z$ is equal
to the sum of two components:\ (1) the cost of a sequential scan of equivalent postings lists stored as contiguous arrays
and (2) the cost of following all pointers that link together non-contiguous slices between different pools.
The first component is the same for all configurations (give a collection) so we can ignore as a constant.
The number of pointers for a term $t$ with frequency $f_r$ can be computed easily given a particular configuration $Z$,
so we can redefine our cost function as follows:
\begin{equation}
\mathcal{C}_T = \sum_{Q \in \mathcal{Q}}\sum_{q \in Q}{|\textrm{Pointers}(\textrm{Postings}(q); Z)| \times C_p}
\label{equation:cachemisscost}
\end{equation}
\noindent where $C_p$ is the cost of following a pointer and $|\textrm{Pointers}(\cdot)|$ is the number of
pointers needed in a particular postings list given a configuration $Z$. The number of
pointers can be easily estimated given the step function $\mathcal{M}$ defined in
Section~\ref{section:memorycost}. Thus, assuming we have an estimate of
the distribution of $|\textrm{Pointers}(\cdot)|$ (from a query log),
we are able to analytically compute a time cost.

What about $C_p$, the cost of following a
particular pointer? Where exactly does this cost come from? Although
all our index structures are held in main memory, latencies can still
vary by orders of magnitude due to the design of cache hierarchies in
modern processor architectures. Reading contiguous blocks of
postings (in a slice) is a very fast operation since (1) neighboring
postings are likely to be on the same cache line, and (2) predictable
memory access when striding postings means that pre-fetching is likely
to occur. 
On the other hand, when posting traversal reaches the end
of a slice, the algorithm needs to follow the pointer to the next slice and begin
reading there---most of the time, this will result in a cache miss,
which will trigger a reference to main memory, which is significantly slower.
Therefore, the cost $C_p$ is dominated by the cost of a
cache miss. However, since we model $C_p$ as a constant, it is
not necessary to estimate its actual value---therefore, our analytical time costs
are modeled in abstract units of $C_p$.

To summarize, we can analytically estimate the time cost if we are given a hypothetical postings length
distribution of query terms and the cost of a cache miss using Equation~(\ref{equation:cachemisscost}).
We stress that this model is overly simplistic and does not account
for time spent intersecting postings. Nevertheless, this simplification is acceptable since
we use the analytical model only to guide our experiments on real data, and in our empirical
results we do measure end-to-end query latency.

\begin{figure}[t]
\includegraphics[width=1\linewidth]{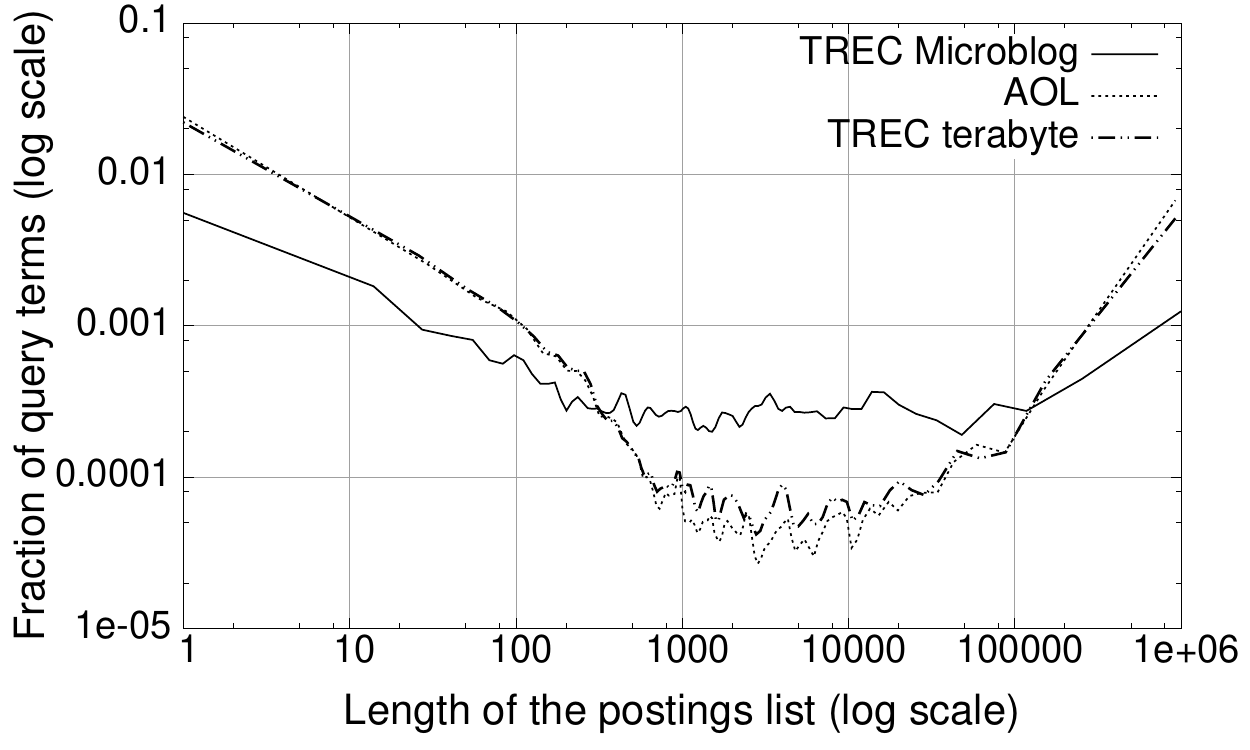}
\vspace{-0.5cm}
\caption{Postings length distribution for various query sets.}
\label{figure:queryCfDist}
\vspace{-0.25cm}
\end{figure}

\begin{figure*}[t]
\includegraphics[width=0.5\linewidth]{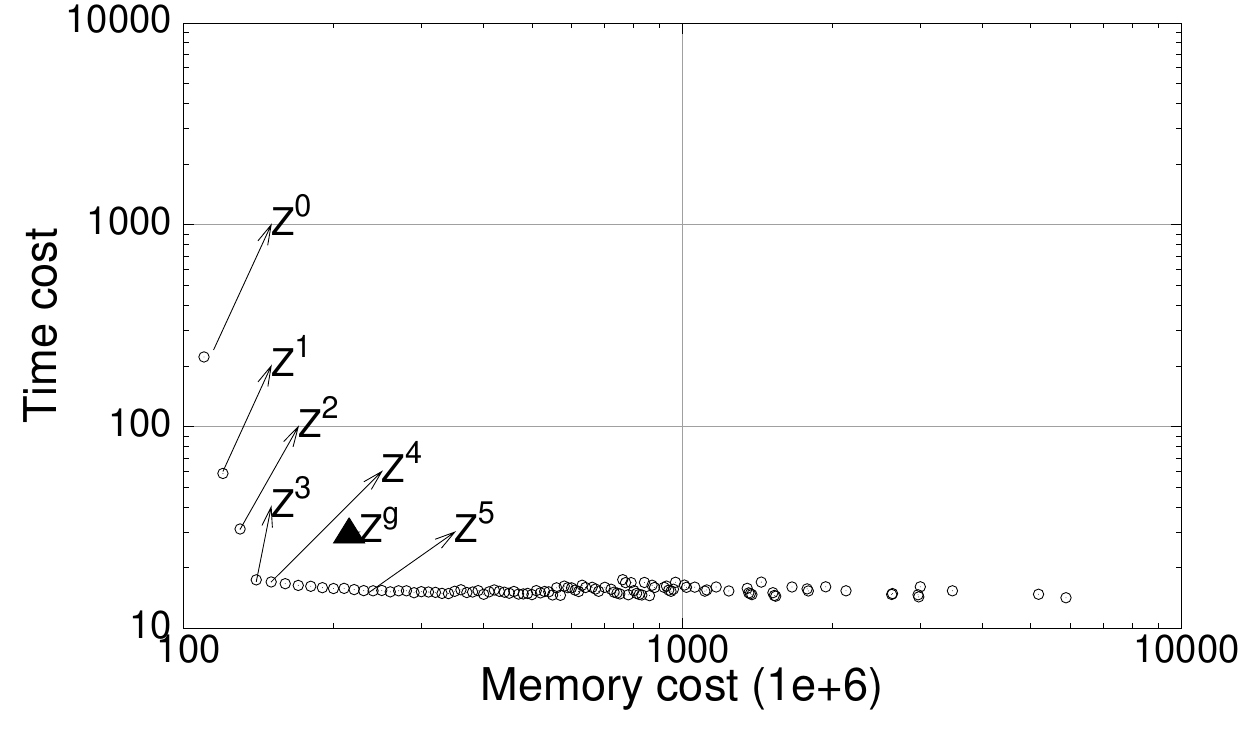}
\includegraphics[width=0.5\linewidth]{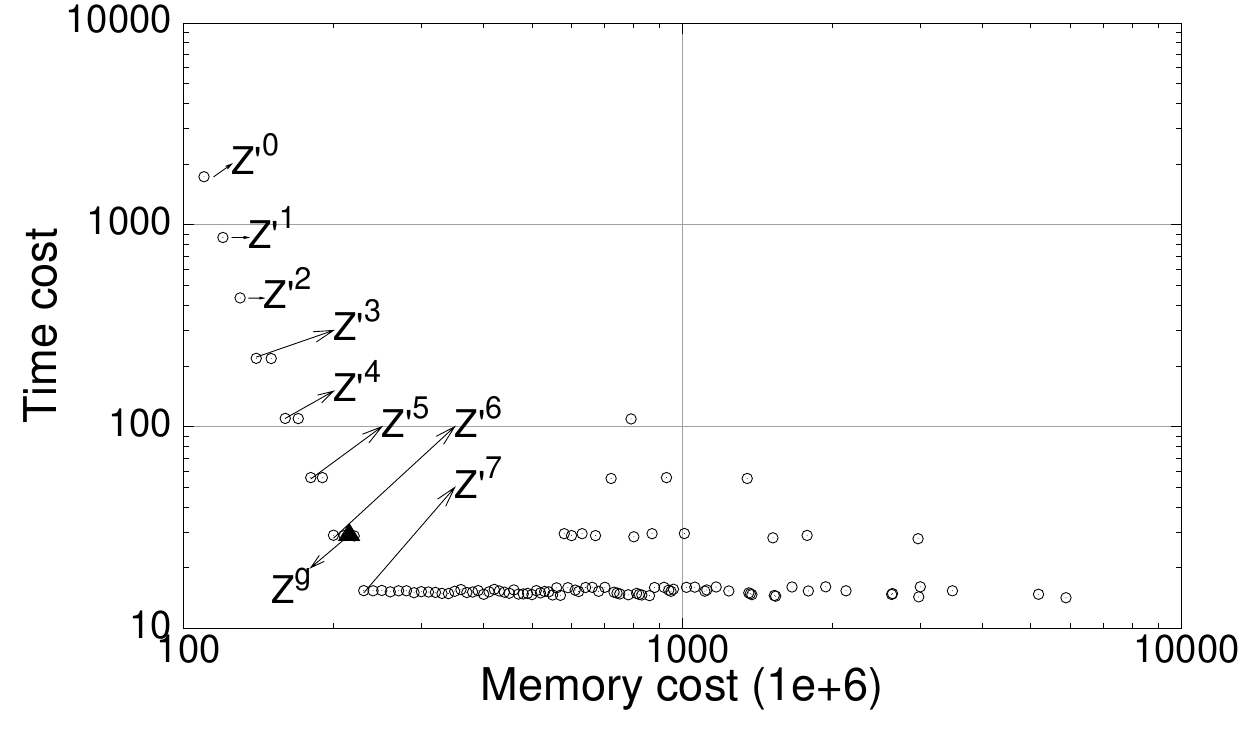}
\vspace{-0.5cm}
\caption{Scatter plot of analytical time cost $\mathcal{C}_T$ versus memory cost $\mathcal{C}_M$, where
each point represents a configuration $Z$. In the right plot, the number of pools is
restricted to 4, whereas in the left plot the number of pools can vary between 4 and 8. Scatter plots shown with
same scale to facilitate comparison.}
\vspace{-0.25cm}
\label{figure:spaceTime}
\end{figure*}

\section{Analytical Results}
\label{section:analyticalResults}

Given a set of configurations $\mathcal{Z} = \{ Z_0, Z_1, ..., Z_m \}$,
we can estimate the memory cost $\mathcal{C}_M$ as well as the simplified time cost
$\mathcal{C}_T$ for any configuration $Z \subseteq \mathcal{Z}$. However, to
complete our model we need to know the total number of terms $N$, size of the
vocabulary $|V|$, and parameter $\alpha$. To determine these values, we divided
the Tweets2011 collection into two equally-sized partitions and used the
first half for parameter estimation; the second half is used in our actual experiments (described later).
We determined $\alpha$ to be $1.0$, and $|V|$ and $N$ to $11 \times 10^6$ and $76 \times 10^6$ respectively.

As explained in Section~\ref{section:timecost}, in order
to estimate the time cost we need the distribution of length of postings for
a set of query terms:\ this is shown for all three
query sets in Figure~\ref{figure:queryCfDist}. This figure
shows that the overall distribution is similar among all query sets.
In particular, the distribution from the AOL and terabyte queries
are nearly identical. Data from the microblog queries give rise to
a similarly shaped distribution, although with less emphasis at the
extremes (both very common and very rare terms). 

Given all these parameters, as well as the set of configurations $\mathcal{Z}$, we estimated the
time cost and the memory cost for each configuration. On a scatter plot
of the time versus memory cost, each configuration $Z \subseteq \mathcal{Z}$
would represent a point: points closer to the origin would be considered
``better'' configurations (faster, less memory). 

Our strategy for exploring the configuration space was to first use our
analytical model to quickly determine the tradeoffs associated with a
large set of configurations, and then from those select a
subset on which to run actual experiments.
We considered slice sizes
between 0 and 12 (inclusive) and pool sizes between 4 and 8 (inclusive)
Another experiment specifically focused on four-pool
configurations (as in the production system). Within these ranges, we computed the memory and time cost
for all possible configurations. Since a scatter plot of all
configurations would not be readable, we grouped the configurations
into equally-sized buckets in terms of memory cost, and
from each bucket, we picked the configuration
that has the smallest time cost. Figure~\ref{figure:spaceTime} shows the scatter plot
constructed in this manner, using the AOL queries for the time cost estimates
(results using other queries look nearly identical, and are not shown for space considerations).
The right plot shows only four-pool configurations; the left plot shows
all pool sizes between 4 and 8 (inclusive).

Based on these figures, we selected a set of candidate configurations
that appear to present good time/cost tradeoffs. As our analytical
models demonstrate, after a certain point the memory costs increase while the time costs
level off, thereby making most of the configurations uninteresting. The
more preferable configurations are those that appear near the origin in
plots in Figure~\ref{figure:spaceTime}. The configurations selected for experimental analysis are
noted.

\section{Using Term History}

There is one additional issue we consider. Given that Earlybird maintains
several index segments in memory (one ``active'', the
rest read-only), it has easy access to historical term statistics from
preceding index segments. It stands to reason that we can take
advantage of this information. Although it seems obvious that such
statistics would help, there are countervailing considerations as
well. We have found that there is a great deal of
``churn'' in tweet content~\cite{Lin_Mishne_ICWSM2012}; for example, approximately 7\% of the top 10,000 terms
(ordered by frequency) from one day are no longer in the top 10,000 on the
next day. This makes sense since discussions on Twitter evolve
quickly in response to breaking news events and idiosyncratic internet
memes. Therefore, using term statistics may not actually help:\ a term
that appeared frequently in the previous index segment may be related
to a news story that is no longer ``hot'', and as a result we might
over-allocate memory and waste space.

To empirically determine how these factors play out on real data, we
experimented with different policies for allocating the first slice
(i.e., instead of always starting from the first pool, choose a pool
with a larger slice size). We refer to this as the Starting Pool (SP)
policy:

\begin{list}{\labelitemi}{\leftmargin=1em}
\vspace{-0.2cm}
\setlength{\itemsep}{-2pt}

\item {\bf SP($z_0$)}: This is the default policy that does not take
  advantage of any term frequency history. Every allocation starts
  from the first memory pool (i.e., $z_0$).

\item {\bf SP($\lceil H(t) \rceil$)}: This policy
starts indexing a term $t$ from the memory pool with
the smallest slice size that is larger than the given
historical frequency $H(t)$, i.e., from the previous index segment. That is, start from pool $p$ if
$2^{z_{p - 1}} < H(t) \leq 2^{z_p}$
or pool $P$ if $2^{z_{P-1}} \leq H(t)$.

\item {\bf SP($\lfloor H(t) \rfloor$)}: According to
this policy, indexing starts from the memory pool with
the largest slice size that is smaller than the given
historical frequency of a term $H(t)$. That is, start from pool
$p$ if $2^{z_{p}} \leq H(t) < 2^{z_{p + 1}}$ or
pool $P$ if $2^{z_{P-1}} \leq H(t)$.

\item {\bf SP($\Lambda(H(t), z_{P - 1})$)}: Based on
this policy, if the frequency of a term $H(t)$ is greater
than or equal to the slice size of the last pool (i.e., $2^{z_{P - 1}}$),
then indexing for that term starts from the last pool.
Otherwise, indexing starts from the default pool, $z_0$.
Function $\Lambda(H(t), z_{P - 1})$ is $z_{P - 1}$ if
$H(t) \geq 2^{z_{P - 1}}$ and $z_0$ otherwise. This basically
divides postings into ``long'' and ``short'', with the last slice
size as the break point.
\vspace{-0.2cm}
\end{list}

\noindent In all of the above policies, when we encounter an out-of-vocabulary
term while indexing, we default to starting from the
first memory pool (i.e., $z_0$).

Using the above schemes, we integrate history into our
allocation policies. Therefore, our experiments explore not
only the impact of different pool configurations, but also the
role of history in improving cost.

\section{Experimental Setup}

To isolate only the effects that we are after, our experiments were
not conducted on the codebase of the live production system, but
rather a separate implementation, which was also
implemented in Java. This allowed us to separate unrelated issues,
such as management of multiple segments, query brokering, and
synchronization of data structures from the core problem of memory allocation.

Experiments were performed on a server running Red Hat Linux, with
dual Intel Xeon ``Westmere'' quad-core processors (E5620 2.4GHz) and
128GB RAM. This particular architecture has a 64KB L1 cache per core,
split between data and instructions; a 256KB L2 cache per core; and a
12MB L3 cache shared by all cores of a single processor. However, all
experiments were run on a single thread.

Our metrics were as follows:\ Evaluation of memory usage is
quantified in terms of memory slots allocated once all
tweets have been indexed (denoted $\mathcal{C}_M^{*}$). Similarly, time costs were measured with
different queries after all the tweets have been indexed. This is a
simplification, since in the production system query
evaluation is interleaved with indexing. However, in production, concurrency is managed
by an elaborate set of memory barriers, which is not germane to the current
study. For our first time metric, we computed the per query
average time to read postings for all query terms in their entirety,
i.e.,

\begin{displaymath}
\mathcal{C}_T^{*} = {1 \over {|\mathcal{Q}|}} \sum_{Q \in \mathcal{Q}}
\sum_{q \in Q}{\textrm{TimeToRead}(\textrm{Postings}(q))}
\end{displaymath}
Unlike estimates from our analytical model $\mathcal{C}_T$,
experimental costs are measured in milliseconds.
In addition, we measured the per query average time to retrieve $k=100$
results in conjunctive query processing mode, i.e., the most recent
100 hits that contain all query terms (we denote this $\mathcal{R}_{100}$). We used a simple linear
merge algorithm to perform postings intersection. Note that although
more effective algorithms are available (e.g., SvS~\cite{Culpepper_Moffat_TOIS2010}), it remains an open question whether they are
suitable for our type of index. Those techniques implicitly assume
contiguous postings lists, since they use variants of binary search to
seek through postings. We felt that to isolate the effects of
different query evaluation algorithms, this was a reasonable choice.

So that we can evaluate the impact of different policies for taking
advantage of term history, we divided the Tweets2011 corpus roughly in
half (chronologically). All experiments were run on the second half,
using statistic from the first half (where appropriate). Note that,
somewhat coincidentally, half of the Tweets2011 corpus corresponds
roughly to the size of the index segments deployed in production,
adding realism to our results.

\begin{table*}[t]
\centering
\small
\setlength{\tabcolsep}{2pt}
\begin{tabular}{|l|r||r|r|r||r|r|r|}
\hline
 & & \multicolumn{3}{|c|}{ postings traversal ($\mathcal{C}_T^{*}$)} & \multicolumn{3}{|c|}{top 100 retrieval ($\mathcal{R}_{100}$)} \\
\cline{3-8}
$Z$ & $\mathcal{C}_M$ & AOL & TB & MB & AOL & TB & MB \\
\hline
\hline
$Z^g = \langle 1,4,7,11 \rangle$ & 90.2m & 1.20 ($\pm$0.02) & 0.86 ($\pm$0.08) & 0.91 ($\pm$0.09) & 2.31 ($\pm$0.01) & 1.58 ($\pm$0.05) & 1.39 ($\pm$0.02) \\
\hline
$Z^0 = \langle 0,1,2,3,4,5,6,8 \rangle$ & 15.9m & 1.33 ($\pm$0.03) & 0.93 ($\pm$0.07) & 0.99 ($\pm$0.06) & 2.02 ($\pm$0.05) & 1.44 ($\pm$0.03) & 1.57 ($\pm$0.02) \\
$Z^1 = \langle 1,2,3,5,6,8,9,10 \rangle$ & 29.1m & 1.21 ($\pm$0.01) & 0.76 ($\pm$0.12) & 0.94 ($\pm$0.01) & 1.90 ($\pm$0.08) & 1.39 ($\pm$0.06) & 1.50 ($\pm$0.03) \\
$Z^2 = \langle 1,3,5,6,8,9,10,11 \rangle$ & 34.9m & 1.19 ($\pm$0.01) & 0.74 ($\pm$0.03) & 0.90 ($\pm$0.01) & 1.89 ($\pm$0.01) & 1.58 ($\pm$0.01) & 1.39 ($\pm$0.02) \\
$Z^3 = \langle 1,3,5,7,8,10,12 \rangle$ & 45.1m & 1.18 ($\pm$0.00) & 0.74 ($\pm$0.02) & 0.91 ($\pm$0.01) & 2.30 ($\pm$0.03) & 1.57 ($\pm$0.01) & 1.69 ($\pm$0.01) \\
$Z^4 = \langle 1,3,6,8,9,11,12 \rangle$ & 49.8m & 1.25 ($\pm$0.01) & 0.74 ($\pm$0.01) & 0.91 ($\pm$0.02) & 2.30 ($\pm$0.01) & 1.57 ($\pm$0.01) & 1.70 ($\pm$0.01) \\
$Z^5 = \langle 2,6,9,12 \rangle$ & 112.1m & 1.23 ($\pm$0.04) & 0.90 ($\pm$0.07) & 0.91 ($\pm$0.01) & 2.30 ($\pm$0.04) & 1.59 ($\pm$0.02) & 1.69 ($\pm$0.03) \\
\hline
$Z'^0 = \langle 1,2,3,5 \rangle$ & 19.7m & 2.71 ($\pm$0.10) & 1.75 ($\pm$0.04) & 1.93 ($\pm$0.09) & 3.14 ($\pm$0.28) & 2.01 ($\pm$0.08) & 2.15 ($\pm$0.14) \\
$Z'^1 = \langle 1,3,5,6 \rangle$ & 24.0m & 1.92 ($\pm$0.04) & 1.20 ($\pm$0.03) & 1.33 ($\pm$0.03) & 2.42 ($\pm$0.13) & 1.67 ($\pm$0.08) & 1.76 ($\pm$0.03) \\
$Z'^2 = \langle 1,3,5,7 \rangle$ & 27.6m & 1.55 ($\pm$0.03) & 1.12 ($\pm$0.17) & 1.11 ($\pm$0.01) & 2.20 ($\pm$0.07) & 1.47 ($\pm$0.01) & 1.69 ($\pm$0.05) \\
$Z'^3 = \langle 1,3,6,8  \rangle$ & 37.3m & 1.36 ($\pm$0.03) & 1.00 ($\pm$0.01) & 1.00 ($\pm$0.01) & 2.04 ($\pm$0.03) & 1.47 ($\pm$0.07) & 1.62 ($\pm$0.10) \\
$Z'^4 = \langle 2,5,7,9  \rangle$ & 59.6m & 1.33 ($\pm$0.13) & 0.89 ($\pm$0.07) & 0.94 ($\pm$0.01) & 1.94 ($\pm$0.01) & 1.36 ($\pm$0.01) & 1.57 ($\pm$0.01) \\
$Z'^5 = \langle 2,5,8,10  \rangle$ & 71.9m & 1.25 ($\pm$0.04) & 0.83 ($\pm$0.07) & 0.92 ($\pm$0.02) & 1.91 ($\pm$0.02) & 1.35 ($\pm$0.01) & 1.58 ($\pm$0.05) \\
$Z'^6 = \langle 2,5,8,11  \rangle$ & 86.4m & 1.25 ($\pm$0.01) & 0.91 ($\pm$0.02) & 0.90 ($\pm$0.01) & 2.34 ($\pm$0.03) & 1.58 ($\pm$0.01) & 1.38 ($\pm$0.02) \\
$Z'^7 = \langle 2,6,9,12  \rangle$ & 112.1m & 1.23 ($\pm$0.04) & 0.90 ($\pm$0.07) & 0.91 ($\pm$0.01) & 2.30 ($\pm$0.04) & 1.59 ($\pm$0.02) & 1.69 ($\pm$0.03) \\
\hline
\end{tabular}
\vspace{-0.1cm}
\caption{Memory cost ($\mathcal{C}_M^{*}$), per query postings traversal time $\mathcal{C}_T^{*}$, and 
per query top {\it k} retrieval time ($\mathcal{R}_k$) for different pool configurations,
using the AOL, terabyte (TB) and microblog (MB) queries on the Tweets2011 corpus (second half).
Time is measured in ms, averaged across 3 trials, with 95\% confidence intervals.
\label{table:results}}
\end{table*}

\section{Experimental Results}

\subsection{Pool Configurations}

Table~\ref{table:results} reports experimental results evaluating
different pool configurations, showing memory cost
($\mathcal{C}_M^{*}$), per query postings traversal time
$\mathcal{C}_T^{*}$, and per query top {\it k} document retrieval time
($\mathcal{R}_k$).  In all cases,
time is measured in milliseconds, and results are averaged across 3
trials, reported with 95\% confidence intervals.  We report results
using the AOL, TREC terabyte (TB) and microblog (MB) queries in
separate columns. The first row of the table shows our production
configuration; the second ``block'' shows select configurations with
the number of pools between 4 and 8 (inclusive); the third ``block''
restricts consideration to 4 pool configurations (as in production).
In all cases we did not take term history into account, i.e., postings
allocation began in the first pool, which corresponds to SP($z_0$).

When considering the 4 pool configurations, analytical modeling
suggests that our production configuration $Z^g$ balances memory and
time quite well (see Figure~\ref{figure:spaceTime}, right). This is
indeed confirmed by our experimental results. Although during the
original implementation of Earlybird no rigorous evaluations
along these lines were conducted, the developers nevertheless honed in on a good point in the
solution space. For example, $Z^{r4}$ and $Z^{r5}$ yield smaller
footprints, and perhaps suggest faster query evaluation, but the
results are inconclusive:\ no significant difference on
$\mathcal{C}_T^{*}$; significantly better for two sets of queries
on $\mathcal{R}_{100}$ but significantly worse for the third. Based on
our results, it doesn't appear possible to significantly speed up
query evaluation, regardless of configuration. On the other hand, it
is possible to dramatically decrease memory consumption by sacrificing
speed, e.g., $Z^{r0}$ (as predicted by our analytical model).

Turning to configurations involving between 4 and 8 pools, we see
opportunities to improve over the current production configuration.
Configuration $Z^{2}$, for example, yields a substantially smaller
memory footprint, while not slowing down query evaluation. However,
the cost is more complex code to manage 4 versus 8 pools (of course,
not modeled in our study). Nevertheless, these experiments point to
possible future improvements in our production codebase.

Note that in this discussion, we avoided use of the term ``optimal'',
since that assumes a single objective metric for combining time and
space in a sensical manner. 
Judgments on the relative merits of memory and speed
must be made with respect to an organization's resources, machine
specifications, etc. For example, we can certainly imagine a case
where $Z^{r0}$ is a good setting---e.g., for academic researchers, where
resources are more constrained and latency demands are perhaps not as
high. Therefore, throughout this paper, we have presented all results
in terms of a memory/speed tradeoff. Any additional attempts to
simplify would be not justified by real-world constraints.

Overall, we find that the predictions made by our analytical model
($\mathcal{C}_M$ and $\mathcal{C}_T$) match the empirical results
quite reasonably ($\mathcal{C}_M^{*}$ and $\mathcal{C}_T^{*}$):\ not
in terms of actual physical quantities, of course, but in terms of capturing the tradeoff
between memory and speed. As we proceed from $Z^{0}$ to $Z^{5}$, and
from $Z^{r0}$ to $Z^{r7}$, memory consumption increases while time
trends downward. However, the overall time differences are not as
large as Figure~\ref{figure:spaceTime} would suggest (i.e., the
vertical axes in the scatter plots are exaggerated).
We note that time estimates produced by our analytical model are in
terms of abstract $\mathcal{C}_p$ units (cost of referencing
non-contiguous postings), not physical time. This congruence
between analytical and experimental results justifies the assumptions
made in our model, and validates the use of analytical estimates to
quickly explore the large configuration space (which is too
large to experimentally explore).
On the other hand, the match between our analytical time cost
$\mathcal{C}_T$ and top 100 retrieval time $\mathcal{R}_{100}$ is not
as good---to be expected, since top $k$ retrieval involves
postings intersection, which is difficult to model analytically.
This points to the limitations of our approach and the need to
perform experiments on real data.

\begin{table*}[t]
\centering
\small
\setlength{\tabcolsep}{2pt}
\begin{tabular}{|l|l|r||r|r|r||r|r|r|}
\hline
& & & \multicolumn{3}{|c|}{ postings traversal ($\mathcal{C}_T^{*}$)} & \multicolumn{3}{|c|}{top 100 retrieval ($\mathcal{R}_{100}$)} \\
\cline{4-9}
$Z$ & SP Policy & $\mathcal{C}_M^{*}$ & AOL & TB & MB & AOL & TB & MB \\
\hline
\hline
\multirow{4}{0.1cm}{\rotatebox{90}{$Z^g$}}
& SP($z_0$) & 90.2m & 1.20 ($\pm$0.02) & 0.86 ($\pm$0.08) & 0.91 ($\pm$0.09) & 2.31 ($\pm$0.01) & 1.58 ($\pm$0.05) & 1.39 ($\pm$0.02) \\
& SP($\lceil H(t) \rceil$) & 104.5m & 1.17 ($\pm$0.00) & 0.74 ($\pm$0.00) & 0.90 ($\pm$0.02) & 2.23 ($\pm$0.01) & 1.44 ($\pm$0.08) & 1.39 ($\pm$0.01) \\
& SP($\lfloor H(t) \rfloor$) & 94.8m & 1.18 ($\pm$0.01) & 0.76 ($\pm$0.01) & 0.92 ($\pm$0.01) & 2.23 ($\pm$0.02) & 1.49 ($\pm$0.07) & 1.39 ($\pm$0.01) \\
& SP($\Lambda(H(t), z_{P - 1}$) & 90.4m & 1.17 ($\pm$0.00) & 0.74 ($\pm$0.01) & 0.91 ($\pm$0.01) & 2.23 ($\pm$0.01) & 1.49 ($\pm$0.06) & 1.41 ($\pm$0.01) \\
\hline
\multirow{4}{0.1cm}{\rotatebox{90}{$Z^2$}}
& SP($z_0$) & 34.9m  & 1.19 ($\pm$0.01) & 0.74 ($\pm$0.03) & 0.90 ($\pm$0.01) & 1.89 ($\pm$0.01) & 1.58 ($\pm$0.01) & 1.39 ($\pm$0.02) \\
& SP($\lceil H(t) \rceil$) & 40.8m & 1.18 ($\pm$0.03) & 0.74 ($\pm$0.02) & 0.92 ($\pm$0.02) & 2.26 ($\pm$0.04) & 1.45 ($\pm$0.09) & 1.39 ($\pm$0.01) \\
& SP($\lfloor H(t) \rfloor$) & 43.2m & 1.16 ($\pm$0.01) & 0.73 ($\pm$0.00) & 0.90 ($\pm$0.02) & 2.26 ($\pm$0.04) & 1.48 ($\pm$0.01) & 1.39 ($\pm$0.01)\\
& SP($\Lambda(H(t), z_{P - 1}$) & 35.0m & 1.17 ($\pm$0.01) & 0.74 ($\pm$0.01) & 0.91 ($\pm$0.02) & 2.25 ($\pm$0.04) & 1.50 ($\pm$0.02) & 1.39 ($\pm$0.01) \\
\hline
\multirow{4}{0.1cm}{\rotatebox{90}{$Z'^5$}}
& SP($z_0$) & 71.9m & 1.25 ($\pm$0.04) & 0.83 ($\pm$0.07) & 0.92 ($\pm$0.02) & 1.91 ($\pm$0.02) & 1.35 ($\pm$0.01) & 1.58 ($\pm$0.05) \\
& SP($\lceil H(t) \rceil$) & 77.7m & 1.22 ($\pm$0.03) & 0.75 ($\pm$0.01) & 0.92 ($\pm$0.01) & 1.90 ($\pm$0.01) & 1.35 ($\pm$0.07) & 1.51 ($\pm$0.02)\\
& SP($\lfloor H(t) \rfloor$) & 73.9m & 1.21 ($\pm$0.03) & 0.77 ($\pm$0.02) & 0.92 ($\pm$0.02) & 1.90 ($\pm$0.01) & 1.35 ($\pm$0.01) & 1.50 ($\pm$0.02) \\
& SP($\Lambda(H(t), z_{P - 1}$) & 72.2m & 1.20 ($\pm$0.00) & 0.75 ($\pm$0.00) & 0.92 ($\pm$0.01) & 1.90 ($\pm$0.00) & 1.35 ($\pm$0.01) & 1.52 ($\pm$0.02)\\
\hline
\end{tabular}
\vspace{-0.1cm}
\caption{Effect of history-based Starting Pool policies. Results are organized the same manner as in Table~\ref{table:results}.
\vspace{-0.3cm}
\label{table:results:historyBased}}
\end{table*}

\subsection{Starting Pool Policies}

In our second set of experiments, we investigated the impact of
Starting Pool policies. As previously described, we divided the
Tweets2011 corpus in half, gathered term statistics from the first half,
and performed experiments on the second half. Experiments focused
on particularly interesting pool configurations from the
previous results:\ $Z^2
\langle 1, 3, 5, 6, 8, 9, 10, 11 \rangle$, $Z'^5 \langle 2,
5, 8, 10 \rangle$, and the default production configuration, $Z^g \langle 1, 4, 7,
11 \rangle$.

When taking advantage of historical term statistics, there are
many issues at play. First, we would expect faster query evaluation
since the postings lists are more likely to be contiguous. This
suggests less time overall when traversing all postings
($\mathcal{C}_M^{*}$), although the impact on $\mathcal{R}_{100}$ is unknown
since top 100 retrieval is unlikely to require traversal of all
postings. In terms of space, there are two considerations:\ starting
at larger slices might save memory due to fewer pointers;
on the other hand, if past statistics are not
entirely predictive, memory will be wasted. How these factors balance
out is an empirical question.

Table~\ref{table:results:historyBased} shows results for various
settings on our three sets of queries. Time is measured across 3 trials with
95\% confidence intervals and the table is organized in a similar manner
as Table~\ref{table:results}. Note that SP($z_0$) is equivalent to
using no term statistics, and is exactly the same as in
Table~\ref{table:results} (row duplicated here for convenience).

Results show that in all cases different SP policies waste space (i.e,
result in a larger memory footprint), without a clear convincing gain
in speed. For example, the most aggressive policy SP($\lceil H(t)
\rceil$) is the most wasteful (8--16\% more memory).
Despite the intuitive appeal of using historical term
statistics, there does not seem to be a benefit, at least for the
policies we studied.

\section{Related Work}

The problem of incremental indexing, of course, is not new. However,
the literature generally explores different points in the design
space.
Previous work typically makes the assumption that the inverted lists
(i.e., postings) are too large to fit in memory and therefore the index must
reside on disk. Most algorithm operate by buffering
documents and performing in-memory
inversion~(e.g.,~\cite{Heinz_Zobel_JASIST2003}), up to the capacity of
a memory buffer. After the buffer is exhausted, inverted lists are
flushed to disk; after repeated cycles of this process, we now
face the challenge of how to integrate the in-memory portion of the
index with one or more index segments that have been written to disk. There are
three general strategies. The simplest is to rebuild the
on-disk index in its entirely whenever the in-memory buffer is exhausted. This
strategy is useful as a baseline, but highly inefficient in
practice. The second option is to modify postings in-place on disk
whenever
possible~\cite{Cutting_Pedersen_SIGIR1990,Tomasic_etal_SIGMOD1994,BrownE_etal_VLDB1994}, for
example, by ``eagerly'' allocating empty space at the end of existing
inverted lists for additional
postings. However,
no ``pre-allocation'' heuristic can perfectly predict postings that
have yet to be encountered, so inevitably there is either not enough space or space is wasted. For the
in-place strategy, if insufficient free space is available, to keep
the postings contiguous, the indexer must relocate the entire inverted
list elsewhere, requiring expensive disk seeks for copying the
data. The third strategy avoids expensive random accesses by
merging in-memory and portions of on-disk inverted lists whenever the
memory buffer fills
up~\cite{Buttcher_Clarke_CIKM2005,Lester_etal_2008}:\ index merging
takes advantage of the good bandwidth of disk reads and writes. In particular,
Lester et al.~\cite{Lester_etal_2008} advocate a geometric
partitioning and hierarchical merging strategy that limits the number
of outstanding partitions, similar to~\cite{Buttcher_Clarke_CIKM2005}.

One challenge of all three strategies described above is the handling
of concurrent queries while in-memory and on-disk indexes are being
processed. No matter what strategy, the operations will take a
non-trivial amount of time, during which an operational system
must continue serving queries efficiently.
Many of the papers cited above do not discuss
concurrent query evaluation. In contrast, this is an important aspect
of our work in building a production system (although not specifically
the focus of this paper).

In the buffer-and-flush approach, Margaritis and
Anastasiadis~\cite{Margaritis_Anastasiadis_CIKM2009} present an
interesting alternative beyond the three strategies discussed
above. They make a slightly different design choice:\ when the
in-memory buffer reaches capacity, instead of flushing the {\it
  entire} in-memory index, they choose to flush only a portion of the
term space (a contiguous range of terms based on lexicographic sort order),
performing a merge with the corresponding on-disk portions
of the inverted lists. The advantage of this is that it does not lead
to a proliferation of index segments, compared to the work of Lester et
al.~\cite{Lester_etal_2008}.

Other than the obvious difference of in-memory vs.\ on-disk storage of
the index, there is another more subtle point that distinguishes
previous work from the Earlybird design. The approaches above
generally try to keep postings lists contiguous---and for good reason,
since disk seeks are expensive. There is, however, substantial cost in
maintaining contiguity in terms of disk operations that are needed at
index time. In contrast, since
Earlybird index structures are in main memory, we found it acceptable
for postings to be discontiguous. While it is true that traversing
non-contiguous postings in memory results in cache misses, the cost of
a cache miss is less in relative terms than a disk
seek. Discontiguous inverted lists allow us to implement a zero-copy
approach to indexing---once postings are written, we never need to
copy them. In a managed memory environment such as the JVM, this
leads to far less pressure on the garbage collector, since buffer
copying yields garbage objects.

In another work, Lempel et al.~\cite{Lempel_etal_CIKM2007} eschew
inverted indexes completely and incrementally build document-centered
representations, from which postings list are dynamically constructed
and cached only in response to queries. The assumption is that more
``heavyweight'' index processes will run periodically (e.g., every 30
minutes), so that all other data structures can be considered
transient. Although this design appears to be justified for the
particular search environment explored (corporate intranet), these
assumptions do not appear to be workable for our setting.

Another interesting point in the design space is represented by
Google's Percolator
architecture~\cite{PengDaniel_Dabek_etal_OSDI2010}, which is built on
top of Bigtable~\cite{ChangFay_etal_OSDI2006}---a distributed,
multi-dimensional sparse sorted map based log-structured merge trees.
Percolator supports incremental data processing through {\it
  observers}, similar to database triggers, which provide cross-row
transactions, whereas Bigtable only supports single-row
transactions. This architecture represents a very different design
from our system, which makes a fair comparison difficult.  The
performance figures reported by the authors suggest that Earlybird is
much faster in indexing, but in fairness, this is
an apples-to-oranges comparison. Percolator was designed to encompass
the entire webpage ingestion pipeline, handling not only indexing but
other aspects of document processing as well---whereas Earlybird is
highly specialized for building in-memory inverted indexes.

Finally, a few notes about our strategy for allocating postings slices
from fixed-size pools:\ there are some similarities we can point to in
previous work, but some important differences as well. With the
in-place update strategy where extra space for postings is
pre-allocated, it is not much of a stretch to implement fixed block
sizes that are powers of two. Brown et al.~\cite{BrownE_etal_VLDB1994}
allocate space for on-disk postings in sizes of 16, 32, 64, 128,
$\ldots$ 8192. However, a few important differences (beyond in-memory
vs. on-disk):\ Brown et al.\ copy postings each time a new block is
allocated to preserve contiguity, whereas we don't. In addition, the
paper leaves open the method by which those blocks are
allocated---whereas we describe a specific implementation based on
fixed slice sizes in large pools (supporting efficient memory
allocation, compact pointer addressing, etc.).

Tracing the lineage of various storage allocation mechanisms further
back in time, we would arrive at a rich literature on general-purposes
memory allocation for heap-based languages (e.g., {\tt malloc} in
C). According to the taxonomy of Wilson et
al.~\cite{WilsonPaul_etal_1995}, Earlybird's allocation strategy would
be an example of segregated free lists, an approach that dates back to
the 1960s. Of course, since we're allocating memory for the very
specific purpose of storing postings, we can accomplish the task much
more efficiently since there are much tighter constraints, e.g., no
memory fragmentation, fixed sizes, etc. Nevertheless, it would be
fair to think of our work as a highly-specialized variant of general
purpose memory allocators for heap-based languages.

\section{Future Work and Conclusion}

Although the problem of online indexing is not new, we
explore a part of the design space that makes fundamentally different
assumptions compared to previous work:\ we consider index structures
that are completely in memory and applications that have much tighter
index latency requirements. There are many
challenges for such applications, and we examined in depth one
particular issue---dynamic postings allocation---within a general
framework for incremental indexing. Our results are interesting in and
of themselves, but we hope to achieve the broader goal of bringing
real-time search problems to the attention of the research community.
Hopefully, this will spur more work in this area.

\section{Acknowledgments} 

\noindent This work has been supported in part by NSF under awards IIS-0916043,
IIS-1144034, and IIS-1218043.  Any opinions, findings, or conclusions
are the authors' and do not necessarily reflect those of the sponsor.
The first author's deepest gratitude goes to Katherine, for her
invaluable encouragement and wholehearted support.  The second author
is grateful to Esther and Kiri for their loving support and dedicates
this work to Joshua and Jacob.

%\bibliography{malloc}
%\bibliographystyle{abbrv}

\end{document}